\documentstyle[aaspptwo]{article}
\tighten

\newcommand{\beq}{\begin{equation}}
\newcommand{\eeq}{\end{equation}}
\newcommand{\beqa}{\begin{eqnarray}}
\newcommand{\eeqa}{\end{eqnarray}}

\newcommand{\ea}{{\it et al. }}

\newcommand{\kmsmpc}{\hbox{$ {\rm km}\, {\rm s}^{-1}\, {\rm Mpc}^{-1}$ }}
\newcommand{\kms}{\hbox{$ \rm km\, s^{-1}$ }}

\begin{document}

\title{A SUPERGIANT SUPERNOVA-BLOWN BUBBLE IN THE SPIRAL GALAXY NGC
1620\footnote{based on observations made at CTIO,  operated by Associated
Universities for Research in Astronomy, Inc. under contract with the
National Science Foundation.} }

\author{
J. Patricia Vader\altaffilmark{2,3} and
Brian Chaboyer\altaffilmark{4,5} }

\altaffiltext{2}{Space Telescope Science Institute,
3700 San Martin Drive,
Baltimore, MD 21218}

\altaffiltext{3}{present address: Les Jardins Etoil\'{e}s,
3001 Dwight Way, Berkeley, CA 94704 e-mail: vader@bkyast.berkeley.edu}

\altaffiltext{4}{Canadian Institute for Theoretical Astrophysics,
60 St.~George St.,Toronto, Ontario, Canada M5S 1A7 \hfill\\
e-mail: chaboyer@cita.utoronto.ca }

\altaffiltext{5}{Department of Astronomy, Yale University,
Box 208101, New Haven, CT 06520-8101}

\begin{abstract}
We present UBR and H$\alpha$ imaging of NGC
1620, a highly inclined spiral galaxy that contains a large scale,
arc-like feature of radius 3 kpc in its outer disk at a distance of
$\sim$ 11 kpc from the center. What is
unusual about this arc-like feature is its stellar nature and the
presence of a luminous star cluster at its center. The arc is
fragmented into HII region complexes and OB star clusters and shows
two kinks in optical continuum light. It spans an angle of 220$^{\circ}$ on
our U image and a full, though fragmented, circle on an unsharp
masked R image. It is centered on a young star cluster that is the
most luminous clump in blue optical continuum light besides the
nucleus of the galaxy. This central star cluster has UBR colors and a
surface brightness similar to those of other HII regions, but is a
relatively weak H$\alpha$ emitter. It consists of
at least three unresolved condensations in optical continuum light.
Its location at the center of the arc and its prominence within the
galaxy suggests that it has been the site of several generations of
supernova explosions that swept up the surrounding gas into a
supershell. When it attained a radius of $0.5-1$ kpc, this shell became
gravitationally unstable and formed the stars which now delineate
the arc. The constraints imposed by the survival of the expanding arc
against random stellar motions and the age of the stars in the arc
yield a required energy input by a minimum of 400 and a maximum
of 6500 supernovae. In this scenario the asymmetry in surface
brightness of the arc reflects the radial gradient of the gas density in
the disk of NGC 1620, while the kinks reflect inhomogeneities in the
original gas distribution with respect to the central star cluster. The
supernova superbubble formed at least $5 \times 10^7$ yr ago so that, unless
supernova explosions continued after the onset of star formation in
the expanding shell, the bubble interior has cooled and the
corresponding hole in the HI distribution no longer exists.
An unsharp masked image reveals the presence of a central bar in
NGC 1620 which we therefore reclassify as an SBbc galaxy.
\end{abstract}

\begin{center}
{\large to appear in   {\it The Astrophysical Journal\/}}
\end{center}
\vspace*{-22cm}
\hspace*{14cm}
CITA--94--53

\section{Introduction}
A shell-like feature with a radius of 3 kpc and centered on a bright
condensation was discovered in the disk of the spiral galaxy NGC
1620 on a blue continuum image (Chaboyer and Vader 1991;
hereafter CV91). The high degree of apparent circular symmetry of
this feature, which is embedded in a highly inclined galactic disk,
suggested that it is a nearly spherically symmetric shell. CV91
interpreted it as a supergiant supernova-blown bubble that formed
$\sim$ $10^8$ years ago as the result of at least
150 supernova explosions. In this scenario the shell of expanding gas
became gravitationally unstable and formed stars which should now
be visible predominantly as stars of type later than B3. We report
here follow-up observations to test this hypothesis. We have
obtained UBR and H$\alpha$ CCD imaging, as well
as low S/N optical spectra of the star cluster at the center of the arc
and of the nucleus of NGC 1620.

NGC 1620 is a highly inclined Sbc galaxy of absolute blue magnitude
$-21.0$, with normal optical and far-infrared properties (Rubin, Ford,
\& Thonnard 1978; CV91) and a typical neutral hydrogen content
(Shostak 1978).  We adopt a distance of 45.7 Mpc based on the
recession velocity found by Rubin et al. (1978) and $H_o =
75~\kmsmpc$. This yields a linear scale of 222 pc per arcsecond.  The
observations and data reduction are presented in Section 2. The
analysis and interpretation of the data are given in Sections 3 and 4,
respectively. Conclusions are summarized in Section 5.

\section{Observations and Data Reduction}

\subsection{Broad Band Optical Imaging}
CCD imaging of NGC 1620 was obtained with the CTIO 0.9m telescope
in November 1990. The detector was a $800 \times 800$ pixel TI Chip,
binned to $2 \times 2$ pixels, yielding $400 \times 400$ pixel
images with a scale of
0.494\arcsec/pixel. The log of the observations is given in Table 1.
On all U,
B and R frames bad columns run East-West across the southern half
of the galaxy. The U and B frames are shown in Figures 1 and 2,
respectively.
All nights were photometric. Ten to sixteen standard stars taken
from Landolt (1983, 1992) and Graham (1982) were observed per
night. The data reduction was done with the IRAF software package.
The images have been flat-fielded and debiased in the standard way.
The instrumental magnitudes are corrected for atmospheric
extinction and converted to UBR magnitudes on the Cousins system.
The transformation equations are given in Table 1. The extinction
coefficients used in B and R are taken from Landolt (1983). From
observations of sixteen standard stars we derive an extinction
coefficient in U of 0.87 mag (unit airmass)$^{-1}$, which is the value used
here rather than that given by Landolt (1983).
\begin{table*}[t]
\begin{center}
\begin{minipage}{14.5cm}
\begin{tabular}{lcccl}
\multicolumn{5}{c}{TABLE 1:~~CCD Imaging Observations}\\
\hline\hline
&\multicolumn{1}{c}{Date} &
\multicolumn{1}{c}{Seeing} &
\multicolumn{1}{c}{Exposure} &
\multicolumn{1}{c}{Transformation}\\
\multicolumn{1}{c}{Filter}&
\multicolumn{1}{c}{Nov 1990} &
\multicolumn{1}{c}{FWHM} &
\multicolumn{1}{c}{Time (s)} &
\multicolumn{1}{c}{Equations\footnote{$u$, $b$, $r$ and $m_{\rm inst}$
are the instrumental magnitudes corrected
to 1 second exposures; $X$ is the airmass.}}\\
\hline
B &  18	& 1.5\arcsec & 2700 & $B =1.008 (b - 0.283 X) - 0.37\pm 0.04$\\
R &  18	& 1.3\arcsec & 1800 & $R = 0.987 (r - 0.130 X) - 0.05\pm 0.03$\\
U &  21	& 1.5\arcsec & 3600 & $U = u - 0.87 X + 0.48 (U-B) - 1.22\pm 0.03$\\
6653 \AA, lc & 17& 1.7\arcsec &3600&
$m_\nu = m_{\rm inst} -0.09 X -6.224 \pm.062$\\
6529 \AA, c  & 17& 1.7\arcsec &3600&
$m_\nu  = m_{\rm inst} -0.09 X -6.518 \pm .049$\\
\hline
\end{tabular}
\end{minipage}
\end{center}
\end{table*}

The photometry errors consist of three components: the systematic
and rms errors of the zero-point in the photometric transformations
(the `external' error), the random error in aperture measurement and
the error in the adopted mean sky value (the `internal' errors). The
sum of the latter two errors defines the relative errors.  We have
calculated the sum of the latter two errors in intensity, i.e., ADU
counts, as $\Delta I = ( N_{tot}/g/m +(\delta n_{sky} A)^2 )^{1/2}$
with $N_{tot}$ the total number of counts as
measured within aperture $A$, $g$ the gain (1.89 e-/ADU), $n_{sky}$ the
mean sky counts per pixel and $\delta$ the
fractional error in the mean sky value. This gives a signal-to-noise
ratio of $\Delta I/ (N_{tot} -A n_{sky})$. The read noise
is low (5.89 e-) and represents a negligible contribution. The
uncertainty in the mean sky value is 0.3\% in B and R and 0.4\% in U.
For areas with surface brightness much below sky the random error
in magnitude can be approximated by $N_{sky}^{1/2}/N_{object}$, while a
relative error $\delta$ in the mean sky yields a
magnitude error of $\delta N_{sky}/N_{object}$.

\subsection{H$\alpha$ imaging}

Two filters were selected for H$\alpha$ imaging: a filter of central
wavelength 6653 \AA ~and a FWHM of 76 \AA, centered on the H$\alpha$
emission line at the redshift of NGC 1620, and an adjacent filter of
central wavelength 6529 \AA ~and a FWHM of 75 \AA ~which was used to
measure the continuum emission in the vicinity of the H$\alpha$ line (Table
1). The line + continuum image suffered from poor tracking, showing
east-west elongated star images with an ellipticity of around 0.1. The
line image was smoothed with an elliptical gaussian to match the PSF
of the poorer-seeing, distorted line + continuum image to form the
final line image. Four standards from Stone and Baldwin (1983) were
observed with both filters. The transformation equations are given in
Table 1, with $m_{\nu}$ the monochromatic magnitude at  6653 and 6529
\AA ~and the extinction coefficient taken from Stone and Baldwin
(1983). We used these results to scale and subtract the `continuum'
frame at 6529 \AA ~from the `line + continuum' frame at 6653 \AA ~(of
similar airmass X) according to
\beq
I({\rm H}\alpha) = I_c - {\rm dex}\,[-0.4(C_c - C_{lc}) ] I_c,
\eeq
with $I$ the intensities, $C$ the zero-points in the transformation
equations (Table 1), and subscripts $c$ and $lc$ referring to the
`continuum' and `line + continuum' frames, respectively. The resulting
H$\alpha$ image is shown in Figure 3a. The stars on the original frames have
been fully removed as expected. The galaxy was offset with respect
to the center of the chip to capture the totality of the arc-like feature
which extends well beyond the disk of NGC 1620. It is positioned
well away from the bad columns of the CCD chip.
Monochromatic magnitudes at 6653 \AA ~are derived from photometry
on this image using the corresponding equation in Table 1. These are
converted to monochromatic fluxes following the prescription of
Baldwin and Stone (1983) and multiplied by the effective filter
width, $\Delta\lambda = 68$ \AA, to obtain H$\alpha$ emission line
fluxes according to $F({\rm H}\alpha) =f\nu c \Delta\lambda/\lambda^2$.
The photometry errors in H$\alpha$ are calculated in the same
way as above, with `sky' replaced by the scaled H$\alpha$  continuum. The
uncertainty in the mean `sky' value is about 1\%.

\subsection{Spectroscopy}
Spectra of the nucleus of NGC 1620 and the central star cluster
within the arc were obtained with the 2D Frutti detector at the CTIO
1m telescope on 24 November 1990. The instrumental setup
corresponded to a slit width of 2\arcsec ~and an effective resolution of
16.5\AA ~FWHM, yielding a nominal accuracy in velocity centroid of
$\sim 100~\kms$. Total exposure times were
1880s and 4200s for the nucleus and the star cluster, respectively.
Observations of two standard stars from Stone and Baldwin (1983)
were used to flux calibrate the spectra. The flat-fielded and
calibrated spectra are shown in Figure 4. The 4000 \AA ~break, the H
and K lines in absorption and H$\alpha$ emission are present in both
spectra. The line features yield average heliocentric velocities of
$3464 \pm 114~\kms$ for the nucleus, in
good agreement with the value of $3496 \pm 10 ~\kms$
found by Rubin et al. (1978), and $3616 \pm 190 ~\kms$
for the star cluster. The spectrum of the star
cluster is much bluer than that of the nucleus. It also shows weak
H$\beta$ in both absorption and emission, which
is indicative of the presence of both ionized gas and young stars.

\section{Data Analysis}
The almost circular arc that stands out on the CCD B-band image (of
seeing 3.3\arcsec ~FWHM) shown in CV91 appears as a less regular and
more clumpy feature on the deeper, better seeing (1.5\arcsec ~FWHM) B
image shown here in Figure 1. The arc and the bright condensation at
its center are most prominent on our U frame (Fig. 2), indicating that
it consists predominantly of OB stars. It spans $\sim 220\deg$
in the disk region of the
galaxy, shows two noticeable kinks, and is broken up in distinct
condensations. Our H$\alpha$ image is less deep
than our U image, which is presumably the reason that many of the
U-bright condensations in Figure 2 are not visible in H$\alpha$.
On the B (Fig. 1) and R (not shown) images which
sample somewhat older stellar populations the arc is a more diffuse
feature.  The central star cluster is irregular in shape and seems to
have at least three components. This is most evident on our R image
which has the best seeing. A contour plot of the central
$9.4\arcsec \times 9.4\arcsec$
region of the clump (Fig. 5) shows a dominant condensation, a fainter
one to the S-W and a more diffuse and redder (see U-R color image
in Fig. 6) extension to the W-N. The cluster lies South-East from the
major axis of NGC 1620, with an actual galactocentric distance of 11.1
kpc if we assume it to be located in the mid-plane of the galactic disk.

A sharply delineated, very straight dust lane, with a width of about
1.8\arcsec, runs along the South-East side of the disk of the galaxy,
at the edge of the region dominated by spiral structure, in a
direction slightly tilted with respect to the major axis of the
galaxy. The dust lane is visible on our U, B and R frames but most
prominently so on the B frame (Fig. 1) and the unsharp masked R frame
discussed below (Fig. 10).  From the shape of the spiral pattern and
the direction of rotation (Rubin et al. 1978), we infer that the S-E
side of the disk is the near side as would also be expected from the
location of the dust lane. The fact that the velocity of the star
cluster is larger than that of the nucleus (by $0.7\,\sigma$,
Sect. 2.3) is consistent with the fact that the South-West side of the
disk is receding from us with respect to the nucleus due to
differential rotation.  The overall continuum light distribution of
the galaxy shows no other perturbations or large scale asymmetries.

\subsection{The brightest H$\alpha$ knots}
Bright knots stand out on the H$\alpha$ line flux
image of the southern two thirds of N1620 shown in Figure 3a. We
have marked the most prominent H$\alpha$ knots
on the overlay for this image shown in Figure 3b (this figure can also
be used as an overlay for Figs. 1, 2, 6 and 10). We can roughly define
a circular ring of radius 14\arcsec ~outlined by seven distinct
H$\alpha$-emitting regions, labeled from 1 to 7 on Figure 3b, and
centered on the position of the star cluster as seen in continuum
light. The position of the star cluster has been chosen as the
geometrical center of its extended light distribution, which lies 3
pixels East of the region of peak intensity in continuum light. We
identify six other bright H$\alpha$ condensations
visible on Figure 3a, labeled A to G in Figure 3b, all of which are
located in the spiral arms of NGC 1620 as outlined by the broad band
images. Regions 7, 1, 2, 3 and 4 are seen to be connected by diffuse
H$\alpha$ emission. Region 6, located near the
eastern edge of the disk, appears isolated and surrounded by diffuse
emission. Weak H$\alpha$ emission can be seen
near the center of the area enclosed by the 14\arcsec ~radius ring. We can
distinguish three low surface brightness peaks --- labeled a, b and c ---
that are
aligned in a South-North direction. The peaks are clearly apparent in
an H$\alpha$ surface brightness profile along a
South-North cut through the center of the ring.  The
center of the 14\arcsec ~radius ring is seen to lie in between peaks a and b.
All the H$\alpha$ knots also stand out on our U
frame, though regions 6 and F are relatively weak. Region 1 appears
isolated. Regions 2 through 5 are part of the continuous though
clumpy arc of enhanced emission. Region 2 is the last bright knot at
the S-W end of the arc, and region 1 is part of a small loop East of the
arc and connecting regions 2 and 3.

We determined position offsets between the H$\alpha$
image and the UBR images using knot D, which is present
on all images, and obtained broadband photometry of the
H$\alpha$ knots. The radial extent (chosen to
include about 90\% of the flux before overlap with a neighboring
region), H$\alpha$  luminosities, optical colors
and positions with respect to the nucleus of the galaxy of the
H$\alpha$ knots are given in Table 2. Magnitudes
and fluxes were corrected for a foreground extinction of $A_B = 0.22 $
for NGC 1620, using the relations
$E(B-V):A_B:E(U-B):E(B-R):A(H\alpha) = 1:4.1:0.78:1.78:2.3$. The relative
or `internal' errors of the magnitudes given in Table 2 are $< 0.02$ mag
in R, $< 0.01$ in B and in the range 0.01 to 0.05 in U, so that the
absolute errors tend to be dominated by the `external' errors due to
the photometric transformations (cf. Sect. 2.1). The net errors in the
H$\alpha$ luminosities given in Table 2 are less
than 10\% for all objects. Figure 7 shows the average H$\alpha$
surface brightness of the individual knots against their
linear extent and a radial profile of differential H$\alpha$
surface brightness of the central star cluster. It is
immediately obvious that the central star cluster is a much weaker
H$\alpha$ emitter than any of the knots. The
plateau in the profile at $r = 10-15\arcsec$ is of course due to the
arc.
\begin{table*}
\begin{center}
\begin{minipage}{17.0cm}
\begin{tabular}{lrrrrrrrrr}
\multicolumn{10}{c}{TABLE 2:~~Properties of HII Regions and the
Central Star Cluster}\\
\hline\hline
\multicolumn{1}{c}{HII} &
\multicolumn{1}{c}{$\Delta x$\footnote{Offset to the East from the
center of the galaxy in cartesian coordinates}} &
\multicolumn{1}{c}{$\Delta y$\footnote{Offset to the North from the
center of the galaxy in cartesian coordinates}} &
\multicolumn{1}{c}{$r$\footnote{Radial distance from the central
star cluster}} &
\multicolumn{1}{c}{aperture} & & & &
\multicolumn{1}{c}{H$\alpha$ flux} &
\multicolumn{1}{c}{L(H$\alpha$)} \\
\multicolumn{1}{c}{region}&
\multicolumn{1}{c}{(\arcsec)} &
\multicolumn{1}{c}{(\arcsec)} &
\multicolumn{1}{c}{(\arcsec)} &
\multicolumn{1}{c}{radius (\arcsec)} &
\multicolumn{1}{c}{B} &
\multicolumn{1}{c}{U -- B} &
\multicolumn{1}{c}{B -- R} &
\multicolumn{1}{c}{($\rm erg\,cm^{-2}\,s^{-1}$)} &
\multicolumn{1}{c}{($\rm erg\,s^{-1}$)}\\
\hline
1& $-21.9$& $-52.9$&10.7& 2.5&  19.62& $ -0.12$ & 1.22 & 4.09E-15 & 1.0E+39\\
2& $-26.4$& $-56.3$&16.2& 3.2&  18.84& $ -0.23$ & 1.10 & 7.16E-15 & 1.8E+39\\
3& $-28.4$& $-48.2$&13.8& 3.2&  18.70& $ -0.25$ & 1.12 & 8.50E-15 & 2.1E+39\\
4& $-25.9$& $-35.8$&14.3& 4.2&  18.10& $  0.11$ & 1.21 & 1.24E-14 & 3.1E+39\\
5& $-8.3 $& $-33.4$&13.3& 2.7&  18.92& $  0.16$ & 1.33 & 2.78E-15 & 6.9E+38\\
6& $-3.4 $& $-47.1$&11.8& 2.7&  19.96& $ -0.08$ & 1.21 & 1.94E-15 & 4.9E+38\\
7& $-15.7$& $-55.3$&10.5& 3.7&  18.83& $ -0.18$ & 1.16 & 5.09E-15 & 1.3E+39\\
A& $-25.6$& $-19.5$&27.5& 4.0&  18.01& $ -0.49$ & 1.04 & 2.07E-14 & 5.2E+39\\
B& $ 5.8 $& $-19.4$&32.9& 3.7&  18.00& $ -0.13$ & 1.22 & 1.49E-14 & 3.7E+39\\
C& $ 9.6 $& $-14.2$&39.3& 2.7&  18.53& $ -0.34$ & 1.09 & 8.28E-15 & 2.1E+39\\
D& $-34.9$& $-52.3$&21.3& 2.7&  19.50& $ -0.63$ & 0.99 & 5.28E-15 & 1.3E+39\\
F& $-15.3$& $  7.2$&52.1& 2.2&  19.52& $  0.19$ & 1.29 & 3.30E-15 & 8.2E+38\\
G& $  0.4$& $-26.1$&24.2& 4.2&  17.99& $  0.15$ & 1.32 & 1.43E-14
 & 3.6E+39\\[3pt]
\multicolumn{6}{l}{Star cluster at the center of the arc}\\
 & $-14.9$ &  $-44.9$ & 0 & 1.0 & 20.70 &$-0.24$ & 1.07 & 7.72E-17 & 1.9E+37\\
 &&&&    2.0&  19.19& $ -0.26$ & 1.01&  5.68E-16 & 1.4E+38\\
 &&&&    3.0&  18.36& $ -0.23$ & 1.00&  1.39E-15 & 3.5E+38\\
 &&&&    4.0&  17.85& $ -0.17$ & 1.01&  2.42E-15 & 6.0E+38\\
 &&&&    4.9&  17.49& $ -0.09$ & 1.05&  2.79E-15 & 7.0E+38\\
 &&&&    5.9&  17.20& $ -0.03$ & 1.08&  3.43E-15 & 8.6E+38\\
 &&&&    6.9&  16.95& $  0.02$ & 1.11&  4.05E-15 & 1.0E+39\\
 &&&&    7.9&  16.73& $  0.07$ & 1.13&  4.79E-15 & 1.2E+39\\
 &&&&    8.9&  16.53& $  0.10$ & 1.14&  5.85E-15 & 1.5E+39\\
 &&&&    9.4&  16.43& $  0.11$ & 1.15&  7.02E-15 & 1.8E+39\\
 &&&&    9.9&  16.34& $  0.12$ & 1.15&  8.72E-15 & 2.2E+39\\
 &&&&   10.9&  16.15& $  0.13$ & 1.16&  1.42E-14 & 3.5E+39\\
 &&&&   11.9&  15.97& $  0.12$ & 1.16&  2.07E-14 & 5.2E+39\\
 &&&&   12.8&  15.80& $  0.12$ & 1.15&  2.70E-14 & 6.8E+39\\
 &&&&   13.8&  15.64& $  0.12$ & 1.15&  3.47E-14 & 8.7E+39\\
 &&&&   14.8&  15.50& $  0.12$ & 1.14&  4.36E-14 & 1.1E+40\\
 &&&&   15.8&  15.37& $  0.12$ & 1.14&  5.19E-14 & 1.3E+40\\
 &&&&   16.8&  15.26& $  0.14$ & 1.15&  5.78E-14 & 1.4E+40\\
 &&&&   17.8&  15.16& $  0.15$ & 1.15&  6.21E-14 & 1.6E+40\\[3pt]
\multicolumn{5}{l}{Nucleus of galaxy}\\
 & 0 & 0 & 47.3 & 4.9 & 15.98 & 0.82 & 1.56\\
\hline
\end{tabular}
\end{minipage}
\end{center}
\end{table*}

A few condensations have H$\alpha$ luminosities
in excess of $3 \times 10^{39}~\rm erg\, s^{-1}$, which would
qualify them as giant HII
regions according to Kennicut and Chu (1988). However, the irregular
shapes, large linear radii (500 to 900 pc) and low surface brightness
(e.g., compare the values in Figure 7 to the limiting outer contour
surface brightness of $0.6\,L_{\sun} \,\rm pc^{-2}$
adopted for HII regions by Hunter
and Gallagher (1985)) of the H$\alpha$-emitting
regions as well as their resolution into clumps on our U image
indicate that they are complexes of HII regions rather than
individual giant HII regions. In this case the luminosities and sizes
are overestimated. On the other hand, if internal extinction is
important the luminosities are underestimated. We derive below an
upper limit to the internal extinction from a UBR color-color diagram
and then consider resolution effects by examining an H$\alpha$
luminosity versus linear size diagram.

\subsection{UBR colors and internal extinction}
In a UBR color-color diagram (Fig. 8) the H$\alpha$
complexes occupy a narrow strip, with apparent B-R
colors typical of F5 to G8 type main-sequence stars and U-B colors
typical of B type main-sequence stars. An internal reddening of
$E(B-V) = 0.75$ would be required for the H$\alpha$
knots to have true colors typical of OB type stars. This is truly an
upper limit, close to the maximum value inferred for better resolved
extragalactic HII regions (Kennicut 1984) and also much larger than
the typical internal extinction of $E(B-V) = 0.16$ expected for a galaxy
of the type and inclination of NGC 1620. In case of a lesser internal
reddening the UBR colors of the H$\alpha$
complexes would reflect a mix of OB type stars superposed on an
older stellar population (cf. Larson and Tinsley 1978). The central
star cluster has UBR colors similar to those of the other HII regions in
NGC 1620. There is no correlation between quantities such as
H$\alpha$ luminosity, H$\alpha$
surface brightness, $\rm L(H\alpha)/L_B$,
position of the knot with respect to the arc, etc... and UBR colors.

The nucleus of NGC 1620 has UBR colors typical of an old stellar
population. On a U-R color image (Fig. 6) the central region of NGC
1620 is distinctly redder than the rest of the galaxy and shows
evidence for some dust patches N-E of the center.

\subsection{H$\alpha$ luminosity versus diameter}
In Figure 9 we have plotted H$\alpha$ luminosity
against diameter. For comparison the results for HII regions in
nearby galaxies (Kennicut 1984) are also shown, with H$\alpha$
luminosities corrected for internal extinction. A least
squares fit to Kennicut's data yields a line of slope 3.5 (solid line in
Fig. 9). Fitting a  line of the same slope to the NGC 1620 data yields a
zero-point shift of 2.25, which corresponds to an increase by a factor
4.4 in diameter at a given luminosity. As NGC 1620 is considerably
more distant than the galaxies in Kennicut's (1984) sample, it is
likely that our bright H$\alpha$ concentrations are a collection of
individual HII regions.  If each H$\alpha$
knot of diameter $D$ and internal extinction-corrected
luminosity $L_c = A L_{\rm obs}$ (with $A$ the extinction factor and
$L_{\rm obs}$ the
observed H$\alpha$ luminosity) consists of $N$
individual HII regions of diameter $d$ and luminosity $l$ with a net
surface filling factor $f$, we have the relations $N\pi d^2 = f\pi D^2$
and $l = A L_{\rm obs}/N$.
Imposing the constraint that the individual HII regions obey the
same luminosity-diameter relation as those in nearby galaxies, we
obtain the relation $(N/f)^{1/2}N^{-1/a} = 4.4A^{-1/a}$, with
$a = 3.5$ the slope of
the luminosity-diameter relation.

We consider two extreme cases: no internal extinction ($A = 1$) and the
maximum internal extinction allowed by the UBR colors ($A = 5$). For
$N = 5$ and $A = 5$ the luminosities remain unchanged, the diameters
decrease by a factor 4.4 and $f = 0.26$. For $N = 5$ and $A = 1$ the
luminosities and diameters decrease by a factor of 5 and 7,
respectively, and $f = 0.1$. In the first case we retain the three giant
HII regions, in the latter case none. We conclude that if we allow for
moderate internal extinction and lack of resolution, the number of
giant HII regions in NGC 1620 is less than 3, which is consistent with
the average frequency of 0.1 and 0.6 giant HII regions per Sb and Sc
galaxy, respectively, found by Kennicut and Chu (1988). On the other
hand, the central star cluster has likely been a giant HII region at the
time of its peak activity.

\subsection{Characteristics of the arc-like feature
and its central star cluster}
An obvious question is to what extent the arc is partial or a complete
circular feature. In an attempt to establish this we have constructed
a U-R color image (Fig. 6) and unsharp masked images (Fig. 10)
which should enhance low surface brightness features. The U-R color
image does bring out the `missing' part of the arc and gives the
impression of a full circular feature that, in contrast to its appearance
on the unsharp masked U, B and R images, is faintest in the direction
toward the center of the galaxy. The ring is everywhere bluer than
its surroundings. The unsharp masked R image is obtained by
dividing the original image by that convolved with a gaussian kernel
of sigma 3\arcsec ~and reduced in intensity by a factor of 0.75. Such an
image effectively reveals faint irregularities concealed by the
(subtracted) global light distribution. In this case it shows a more
diffuse version of the lumpy arc seen against the disk of NGC 1620
on our U image as well as a fainter ridge that completes the circle.
This ridge shows region 6 and two fainter adjacent clumps as local
luminosity enhancements and runs across the dust lane at very low
surface brightness (which is seen much better on the original image
than on the print reproduced here). Some low surface brightness
features are also visible within the circle. The spiral arm containing
regions B, C and G cuts across the circle at lower surface brightness
and seems to end in knot 7. Two faint wisps seem to emanate from
the central star cluster, one to the North which reaches the arc near
knot 5 and one to the South which ends in knot 7. The wisps do not
seem to be part of a spiral arm. Since a supernova bubble event
affects the surrounding gas but not the stars the presence of spiral
arms or other features associated with the older stellar population
within the bubble region is expected. An unsharp masked B image
looks very similar but has less contrast than the unsharp masked R
image because the B image is less deep and the B continuum is a
relatively less strong component. For the same reasons an unsharp
masked U image is even less informative.

\section{Discussion}
CV91 proposed that the giant arc in NGC 1620 is the result of a
supernova-blown bubble whose shell became gravitationally
unstable and formed stars. On the basis of models of superbubbles
(Mac Low and McCray 1988, McCray 1988), they argued that an
energy injection by 7000 supernovae over a period of $7.6 \times 10^6$ yr,
yielding a final velocity of the shell at the time of star formation of
$50~\kms$, accounts best for the current size of the arc.
In the light
of our new observational results, we reconsider the `supergiant
bubble' interpretation of the arc here in some more detail.

\subsection{The supergiant bubble scenario}
The young star cluster at the center of the arc stands out as the
brightest and most extended condensation in continuum light besides
the nucleus of the galaxy (cf. Table 2). While the spiral arms of NGC
1620 are well delineated by bright condensations in continuum light,
this star cluster is obviously not part of a spiral arm as we would
expect it to be if it were a `normal' HII-region complex (for example,
the supergiant HII regions in M101 all obviously lie in spiral arms).
On our U and B images the annulus in between the central cluster
and the arc is remarkably `empty' of any features superposed on the
background continuum. Our unsharp masked R image dramatically
illustrates the prominence of the cluster against a background of
stellar spiral arms and faint wisps which would of course not have
been affected by any relatively recent supernova events.
Because of its unique and isolated location at the center of the arc
and its large luminosity, it seems most likely that the young star
cluster itself is responsible for its unusual environment within the
galaxy. On the basis of its redshift, its diffuseness that is
characteristic of a young star cluster rather than of a galaxy nucleus,
and the stability arguments discussed by CV91, it can be ruled out
that the blue clump is an intruder or an accreting companion of NGC
1620. The most likely interpretation therefore is that the star cluster
has been the site of a large number of supernova explosions that
swept up the surrounding gas into an expanding shell which became
gravitationally unstable and formed stars. The fact that the central
star cluster consists of at least three lumps makes it plausible that a
continuous supernova input over several generations of stars has
occurred.

A sustained energy input would have been extremely favorable for
the formation of a supershell. Such a shell is three-dimensional. If
the central star cluster is located in the mid-plane of the galaxy, the
expanding gaseous shell will remain roughly spherical as long as its
radius does not exceed one vertical scale height of the ambient gas.
At larger radii the shell would expand more rapidly in the vertical
direction, acquiring a prolate shape . The distension of the shell in
the vertical direction would result in a final appearance of the shell
as a ring in the plane of the galaxy with a terminal radius of about
two gas scale heights (McCray 1988). In the case of NGC 1620,
projection effects due to the $\sim 70\deg$
inclination of the disk would result in
an ellipse with axial ratio $\sim$ 3. Our U
frame (Fig. 2) clearly shows that the arc, if elliptic at all, appears
much less elongated than the spiral arms and inner regions of the
disk. The implication is that the arc is a 3-dimensional shell rather
than a ring so that the expanding HI shell must have become
unstable and formed stars before breaking out of the galactic disk.
On the other hand, the arc, with two noticeable kinks (near
H$\alpha$ knot 3 and in between knots 4 and 5)
and hardly visible in continuum light to the South-East, is not
perfectly circular and is distinctly asymmetric in surface brightness.
Given the presence of a strong gradient in optical continuum light
and of faint spiral arms (Fig. 10), we speculate that large-
scale gradients and inhomogeneities in the density of the ambient
gas are most likely responsible for this, e.g., the faintness of the arc
in the direction away from the center of the galaxy is most likely
caused by a negative radial gradient in the gas density while the
kinks may be due to encounters with gas clouds in the nearest spiral
arms.

The dust lane in the disk of NGC 1620 (Figs. 1, 2, 10), at a projected
minimum distance of 7.5\arcsec ~from the center of the star cluster, appears
unperturbed. This implies that the expanding shell must have
formed stars when it had a radius smaller than 1.7 kpc, or about half
its current size. The fact that the arc is seen to run across the dust
lane in continuum light (Fig. 10) is consistent with this scenario.

\subsection{The arc as the stellar remnant of a
super\-nova-blown bubble}
At early times a continuous supernova energy input keeps the
interior of a bubble pressurized and drives the expansion of the shell
until a time $t_1$ at which gravitational instabilities set in, leading to
star formation (cf. McCray 1988).  The time $t_1$ and the corresponding
radius and velocity of the shell can be written as
\beqa
t_1 & \approx & 1.3\times 10^7\,L_{39}^{-1/8}\, n_o^{-1/2}\,
a_s^{5/8}~~{\rm yr}\\
R_1 & \approx & 500\, L_{39}^{1/8}\, n_o^{-1/2}\, a_s^{3/8} ~~{\rm pc}\\
V_s & \approx & 22.5\,L_{39}^{1/4}\, a_s^{-1/4}~~\kms ,
\eeqa
with $L_{39}$ the power input by supernovae in units of
$10^{39} \rm erg\, s^{-1}$, $n_o$
the atomic number density of the ambient gas
---assumed to be uniform--- in
units of $\rm cm^{-3}$ and $a_s$ the magneto-sound speed in the
shell in units of $\kms$. The supernova power input is given by
$L = \epsilon {\rm E_{sn} }$, with $\epsilon$ the supernova
frequency and $\rm E_{sn}$ the energy per supernova, so that $L_{39}$
corresponds to a frequency of one supernova per
$3.2 \times  10^4$ yr for ${\rm E_{sn}} = 10^{51}$ erg.
We adopt a magneto-sound speed of $1~\kms$ in the shell,
as is characteristic of a weak magnetic field.

If the radius $R_1$ at which the shell formed stars did not exceed the
vertical scale height of the gas, the shell has remained roughly
spherically symmetric. For $t > t_1$, the now stellar shell continues
expanding at constant velocity $V_s$ provided that $V_s$ exceeds the
ambient random stellar velocity, i.e., $V_s > 20~\kms$ (expansion
velocities of $\sim 20~\kms$ are typical of 1
kpc large HI supershells in our Galaxy (Heiles 1979)). The radius of
the shell becomes $R(t) =  R_1+ V_s t$. The constraint on $V_s$ requires a
supernova power input $L_{39} > 1$, and, for
$n_o \approx 1\,{\rm cm}^{-3}$,
yields a minimum radius $R_1 \sim 500$ pc and a maximum possible age
of the stars in the
shell of $1.3 \times  10^8$ yr, while $t_1$ (eq. [2]) remains smaller than the
lifetime $\sim 5 \times 10^7$ yr of a type B3 star,
the least massive supernova progenitor, so that one coeval
generation of stars can assure a continuous power supply. The total
number of supernovae required in this case would be $> 400$. On the
basis of their UBR colors, the H$\alpha$ knots in
the shell could consist of a single generation of OB stars or be a
stellar population mix with a wider age range, the uncertainty being
due to the unknown internal extinction. In the extreme case of OB
stars only, the age of the stars in the shell should be less than
$5 \times 10^7$ yrs, which is the
lifetime of the least massive ionizing star, of
type B3. This would require an expansion velocity of $50~\kms$
implying a supernova power of $L_{39}\approx 24$
injected over $8.7 \times 10^6$ yr, or 6500 supernovae. This is comparable to
the number of supernova progenitors in the largest OB associations in
our Galaxy, as estimated from the ionizing radiation of radio HII
regions (Heiles 1990, McKee and Williams 1993). This case yields a
maximum radius of the gaseous shell, $R_1 \approx 750$ pc, which
 is comparable to that of the HI supershells
found in our Galaxy. Larger supernova energy inputs would yield
larger $V_s$ and $R_1$, and a smaller age of the stars in the shell. Because
the gas crossing time of a cavity of radius $R_1$,
$\tau = 5 \times  10^7\, (R_1/500\,{\rm pc})(v_{\rm gas}/10\,\kms)$ yr,
is comparable
to the total time elapsed since the onset of the formation of the
postulated superbubble this cavity would no longer exist as a hole in
the HI distribution of NGC 1620 unless supernova activity has been
sustained at times $t > t_1$.

According to the above superbubble scenario the cluster at the
center of the arc is older by $\sim 10^7$ yr
than the stars and HII regions in the arc. This is consistent with the
fact that the blue clump seems older than those HII regions on the
basis of its weaker H$\alpha$ emission (Table 2).
If the blue clump is a supergiant star cluster or a complex of star
clusters that originally formed within a spiral arm, then the
explosion of a large number of a supernovae must have evacuated
the surrounding region, and the accumulating gas swept up by an
expanding bubble would have run into and distorted the gas trapped
in the nearest spiral arms. The diffuse H$\alpha$
emission near the central star cluster can be due to gas initially
associated with that cluster or newly acquired gas by thermal
evaporation within the hot bubble.  We note that much of the above
discussion depends on the ambient gas density being of order $1~{\rm
cm}^{-1}$, a value which would appear to be rather large for gas
$\sim 500$ pc from the plane.  However, we note that CO measurements
in NGC 891 by Scoville \ea (1993) indicates that the gas in that
galaxy has twice the scale height of the gas in our own Galaxy.  The
existence of the shell may be taken as an indication that
the HI layer is much thicker in NGC 1620 than our Galaxy.

The inside of the bubble seems remarkably smooth and `empty'
except for the underlying continuum light of the older stellar
population, including spiral arms and two faint wisps (Fig. 10). This
suggests that all gas initially present has been swept up or, in case of
denser clouds, thermally evaporated in the hot bubble interior. The
cooling time of the hot bubble interior,
\beq
t_c \approx 3\times 10^7\,L_{39}^{3/11} \,n_o^{-8/11}~~{\rm yr}
\eeq
(Mac Low and McCray 1988), is smaller than the time elapsed since
the formation of the bubble so that, unless supernova explosions
continued after the onset of star formation in the expanding shell,
the bubble interior has cooled and is no longer observable in soft X-
rays.

The swept up mass,
\beq
M \approx 1.3\times 10^7\, R_{500}^3\, n_o ~~M_{\sun}
\eeq
with $R_{500}$ the radius $R_1$ in units of 500 pc, is comparable to the HI
masses found in supershells in the Milky Way and other galaxies (cf.
van der Hulst and Kamphuis 1991), while the typical mass of a
gravitationally unstable fragment
\beq
M \approx 1.3\times 10^5\, L_{39}^{-1/8}\,n_o^{-1/2}\,a_s^{29/8} ~~M_{\sun} ,
\eeq
(McCray 1988) is a hundred times smaller and comparable to the
mass of a large molecular cloud in the Galaxy.

\subsection{The location of the arc and the local structure of the
galactic disk}
Because of the overall remarkably circular shape of the arc, the
distinct asymmetry in surface brightness of the arc in NGC 1620
presumably reflects the radial gradient of the gas density along the
galactic disk rather than a partial break-out of the original shell from
the gaseous disk. If no break-out occurred, then the size of the shell
in its gaseous state has not exceeded a few scale heights of the
gaseous layer (Mac Low et al. 1989). It follows that the gas scale
height at the location of the arc should be of the same $0.5-1$ kpc
order as the values derived above for $R_1$. By analogy with two well-
studied Sb galaxies, our Galaxy and the edge-on spiral NGC 891, this
is quite a reasonable result. The galactocentric distance to the center
of the arc of 11.1 kpc is somewhat larger than that of the solar
neighborhood in our Galaxy. The scale height of the neutral HI gas in
the solar neighborhood is $\sim 200$ pc
(and likely a factor of two large in NGC 891 Scoville \ea 1993)
and
known to increase rapidly with increasing distance, up to $500 - 1000$
pc beyond the solar circle (e.g., Kulkarni et al. 1982). Such an
increase is expected from a decreasing net surface mass density of
the disk to which the gas scale height is inversely proportional (cf.
Vader and de Jong 1981). The surprisingly large size of many
supershells has emphasized the importance of a low-density tail of
HI gas at large heights above the plane of our Galaxy originally
discovered by Shane (1971), and of the extended diffuse ionized
layer of gas with a typical scale height of 1 kpc, discovered by
Reynolds (1990) in our Galaxy and detected in other nearby spiral
galaxies such as NGC 891 (Rand et al. 1990; Dettmar 1990). The
existence of this ionized layer and that of superbubbles, shells,
worms and chimneys (Heiles 1990) are believed to be inter-dependent,
the latter forming the channels through which the
ionizing photons escape confinement within the disk and the gas
acting as a confinement agent up to large vertical distances from the
disk (Mac Low et al. 1989, Norman and Ikeuchi 1989, Norman 1991).
The fact that all supershells in our Galaxy are located beyond the
solar circle offers indirect support of the confining effects and
outwardly increasing scale height of the gaseous layer. No correlation
between HI shells and OB associations or HII regions has been found
in the Galaxy (Heiles 1979, 1984). In contrast, the stellar arc in
N1620 is remarkable because of the very luminous blue star cluster
at its center which must be at its origin. At the location of the bubble,
the disk surface brightness presents a strong gradient in the N-S
direction which presumably reflects the radial gradient in
surface mass density, implying a strong increase outwards of the
gaseous scale height.

The annulus around the central star cluster and within the arc, which
is devoid of any local features above our detection limit except for
faint stellar spiral arms and wisps, has colors $B-R = 1.45$ and
$U-B = 0.2 $ (Table 2) and an R surface brightness of
21.5 mag arcsec$^{-2}$, yielding a B surface brightness corrected
for inclination and
internal extinction of 23.8 mag arcsec$^{-2}$. These values are typical of
the inter-arm regions of the disks of spiral galaxies (Schweizer 1976).
The colors also fit in a sequence of stellar population models with
decreasing star formation rates and age $10^{10}$ yr (Larson and Tinsley
1978) and suggest a mass-to-light ratio of the stellar population of
$M/L_B \sim 2$ to 3.
These results indicate that the arc can be explained as the product of
a supergiant bubble under totally ordinary circumstances in which
the disk of NGC 1620 has a structure typical of the outer regions in
spiral galaxies.

\subsection{NGC 1620 as an ordinary SBbc galaxy}

Except for its giant arc, NGC 1620 appears to be a normal SBbc
galaxy. An inspection of our optical broad-band images shows that
overall light distribution in the disk of NGC 1620, at the
galactocentric distance of the arc and beyond, is very symmetric with
respect to the center of the galaxy. Hence the underlying older stellar
population is not affected by the presence of the clump. A global HI
flux measurement  yields a total HI mass of $1.1 \times  10^{10}$ solar masses
and a hydrogen mass to luminosity ratio $M_H/L_B = 0.35$ (Shostak
1978) that are typical for an Sbc galaxy. We have commented above
upon the number of giant HII regions in NGC 1620 (Sect. 3.3). In the
case of NGC 891, an edge-on Sb galaxy, it is remarkable that, in spite
of the scarcity of giant HII regions ---the
sites of powerful supernova explosions---
in Sb galaxies (Kennicut and Chu 1988), an extended layer of ionized
gas with a scale height of order 1 kpc has been detected in
H$\alpha$ (Rand et al. 1990; Dettmar 1990) as
well as possibly evidence for outflow of gas from the disk through
chimneys (Norman and Ikeuchi 1989). This galaxy must therefore
have a disk-halo connection powered by star formation. NGC 1620
may be another example of such a galaxy. The presence of what is
most likely the remnant of a supergiant HII region that is
responsible for the creation of a superbubble and the observed arc is
then not so surprising. Either NGC 1620, NGC 891 and our Galaxy, all
of which show features believed to be associated with the occurrence
of giant HII regions, have a larger number of giant HII regions than
expected for Sb-Sc galaxies, or the numbers found by Kennicut and
Chu (1988) on the basis of ten Sb and fifteen Sc galaxies are not
representative, or ordinary HII regions are sufficient to sustain a
disk-halo connection.

\section{Conclusions}
Our UBR and H$\alpha$ imaging of NGC 1620
shows that the smooth arc visible on the POSS print and on an earlier
CCD B image (CV91) consists of a young stellar population and
contains several HII region complexes. On a U-R and an unsharp
masked R image, a faint ridge of continuum light is detected on the
East side and completes the arc into a ring. We have established that
the clump at its center is a very luminous star cluster which, given
its relatively weaker H$\alpha$ emission, is
presumably somewhat older than the stars in the arc. Our new data
support the hypothesis of CV91 that the arc is a supernova-induced
supershell that formed stars. We have considered two extreme
possibilities constrained by the survival of the expanding arc against
random stellar motions and the age of the stars in the arc. Requiring
a minimum expansion velocity of $20~\kms$ yields an energy input
from $> 400$ supernovae over a time scale of  $1.3 \times 10^7$ yr, a lower
limit of 500 pc for the radius of the HI shell at the onset of star
formation and a maximum age of the stars in the arc of $1.3 \times 10^8$ yr.
At the other extreme, requiring an age of the stellar arc of less than
$5 \times 10^7$ yr ---the lifetime of the least
massive ionizing star--- yields an
expansion velocity of the arc of $50~\kms$, an energy injection by
6500 supernovae over $8.7 \times 10^6$ yr, and a radius of the HI supershell
of 750 pc. The latter case corresponds to a number of supernova
progenitors that is comparable to that estimated for the largest OB
associations in our Galaxy. This scenario requires that the NGC 1620
has a much thick layer of HI as compared to our own Galaxy.
The maximum radius of the HI shell at the
onset of star formation is less than three times the current size of the
arc. In this case the hole of similar size created in the HI distribution
by the superbubble no longer exists because the time elapsed since
its formation exceeds the gas crossing time. On the other hand,
depending on the efficiency of star formation, an HI enhancement
might still be visible at the location of the arc. Because NGC 1620 is a
not quite edge-on galaxy at a relatively large distance, a study of its
vertical disk structure, e.g., the detection of a thick disk of ionized
gas, is difficult. However, high resolution radio continuum
observations are desirable for a determination of the internal
extinction of the HII regions in the arc which would yield a better
estimate of the ionizing flux and of the average age of the ionizing
stars from extinction-corrected optical colors. It would also be
interesting to try to determine the current expansion velocity of the
arc with, e.g., Fabry-P\'{e}rot observations in H$\alpha$.

The location of the arc in the outer disk of the galaxy, at about 11
kpc from the center, is typical of that of HI supershells in other spiral
galaxies where their existence is favored by gas with large scale
heights that acts as a confining agent. The circular shape of the arc in
the highly inclined disk of NGC 1620 suggests that blow-out during
the gaseous phase has not occurred to a significant degree.
What is unusual about the arc-like feature in N1620 is its stellar
nature and the presence of the luminous star cluster at its center. We
do not know of similar configurations in other galaxies. NGC 1620 is
an otherwise unremarkable SBbc galaxy.

In conclusion, the large scale arc and its central star cluster in NGC
1620 present an interesting test case for current ideas on supergiant
shells and associated features and how they establish a disk-halo
connection.

\acknowledgments
The comments of the anonymous referee improved the presentation of
this paper.  BC gratefully acknowledges the hospitality of the
Space Telescope  Science Institute during several visits.

\appendix

\clearpage

\begin{centering}
\Large \bf FIGURE CAPTIONS \\
\end{centering}

\begin{description}
\item[FIG. 1:]
$125\arcsec \times 125\arcsec$ B image on a linear
intensity scale. North is up and East is to the left.

\item[FIG. 2:]
U image (Nov 21) on a linear
intensity scale. Size and orientation same as in Fig. 1.

\item[FIG. 3:]
 a) H$\alpha$
line emission image on a linear intensity scale. Size and orientation
same as in Fig. 1. The brightest condensations are identified. The
circle of radius 14\arcsec ~indicates the position of the arc.
	 b) An overlay of a) identifying the H$\alpha$
knots (open circles for knots 1 to 6, triangles for knots A
to G), the central star cluster and the nucleus of the galaxy (crosses),
and the same circle as in a). This overlay can also be used for Figs. 1,
2, 6 and 11.

\item[FIG. 4:]
Spectrum of the star cluster at the
center of the arc and of the nucleus of NGC 1620. The vertical lines
identify the CaII K and H lines in absorption and H$\alpha$
in emission. H$\beta$  is seen in both
absorption and emission for the star cluster.

\item[FIG. 5:]
A contour plot of the R intensity
on a linear scale, centered on the blue clump. The intensity of the
contours decreases in steps of 150 counts, from 5277 to 4077
counts/pixel, which corresponds to an observed surface brightness
range (not corrected for extinction) of 20.83 to 21.75 mag arcsec$^{-2}$.
The spatial scale shown is in pixels (0.494\arcsec/pixel).

\item[FIG. 6:]
Color map U-R on a logarithmic
intensity scale. Size and orientation same as in Fig. 1.

\item[FIG. 7:]
Average H$\alpha$ surface brightness (corrected for foreground extinction)
against radius of the knots in NGC 1620 (open circles for knots 1 to 6;
triangles for knots A to G ) and the differential H$\alpha$
surface brightness profile centered on the star cluster at
the center of the arc (dots).  Note that the center of the star
cluster has been defined on the B frame, and that there is strong
off-centered H$\alpha$ emission (Fig. 3), so that the peak H$\alpha$
intensity in the star cluster does not correspond to 0 radius.

\item[FIG. 8:]
UBR color-color diagram, with
colors corrected for foreground extinction. Open circles represent
knots 1 to 6; triangles knots A to G and the two dots represent
circular apertures of 5\arcsec ~radius centered on the blue clump and on the
nucleus of the galaxy (cf. Table 2). The solid curve is defined by
main-sequence stars from type O9.5 to K5, the dotted curve by giant
stars of type G5 to K1 (Johnson 1966). The arrow is a reddening
vector of a length corresponding to $E(B-V) = 0.5$.

\item[FIG. 9:]
H$\alpha$
luminosity (erg s$^{-1}$) corrected for foreground extinction against
diameter (pc) for the H$\alpha$ knots of NGC 1620
listed in Table 2 (open circles for knots 1 to 6; triangles for knots A
to G ). The diamonds represent extragalactic HII regions with
H$\alpha$ luminosity corrected for foreground
and internal extinction (Kennicut 1984; his Table 2). The solid lines
represent linear square fits to the data, with a slope of 3.5,
determined from Kennicut's data. The dotted line shows the location
the HII  regions in NGC 1620 would have if the H$\alpha$
knots were resolved in 4 individual HII regions with a
surface filling factor of 50\% (see text).

\item[FIG. 10:]
An unsharp masked R image
obtained by dividing the original R image by the R image convolved
with a gaussian kernel of sigma 3\arcsec ~and reduced in intensity by a
factor 0.75. The size is $195\arcsec \times  195\arcsec$. The scale,
the orientation and the
circle are the same as in Fig. 3
\end{description}
\end{document}